\documentclass{svproc}
\usepackage{graphicx}
\usepackage{url}

\begin{document}
\mainmatter           

\title{Tree Code Based Neighborhood Algorithms for Discrete Element Methods}
\titlerunning{Tree Code Based Neighborhood Algorithms}  \author{Yuki Watanabe\inst{1} \and Dominik Krengel\inst{2} \and Hans-Georg Matuttis\inst{1}}

\authorrunning{Yuki Watanabe et al.} \tocauthor{Yuki Watanabe, Dominik Krengel, and Hans-Georg Matuttis}
\institute{The University of Electro-Communications, 1-5-1, Chofu, Tokyo, 182-8585,
Japan\\
\email{hg@mce.uec.ac.jp}\\
\and
Tsukuba University, 1-1-1, Tennodai, Tsukuba, 305-8573, Japan\\
\email{krengel.dominik.kb@u.tsukuba.ac.jp}\\
}

\maketitle              

\begin{abstract}
We report our experiences for the development of a neighborhood algorithm implemented via tree-codes to optimize the performance of a discrete element method (DEM) for convex polytopes. Our implementation of the two-dimensional tree code needs $N\log N$, as the does the sort and sweep approach.  For our choice of boundary conditions (a rotating drum) and system sizes (up to several thousand particles), the performance of the tree-code is slightly better, but the algorithm is considerably more complicated than the sort and sweep approach.

\end{abstract}
\section{Introduction}
In the discrete element method (DEM), the most time-consuming part is the computation of the inter-particle forces, followed by the neighborhood algorithm to determine which particles can interact.
Currently used  neighborhood algorithms have various drawbacks which are detailed in the next subsection. As tree data structures allow processing with minimal complexity, we have implemented a neighborhood algorithm based on a tree code. As DEM simulation, we use a two-dimensional ``hard particle, soft contact'' type where the repulsive force is proportional to the particle overlap. Particles are modeled as convex polygons which can be elongated with some shape dispersion. All codes are programmed in MATLAB.

\subsection{Conventional neighborhood algorithms}
A straightforward implementation of long-range forces (every particle interacts with every other particle) for an $N$-particle system needs $1/2  N(N  - 1)$ or  $O(N^2)$ operations. (The factor $1/2$ is due to Newton's principle of action=reaction, so that each pair has to be computed only once.) 
In DEM-simulations with particles which interact only in direct contact, 
the number of operations should be proportional to the number of the particles. Typically,  one has $\approx 1/2 \cdot 6=3 $ in two and  $\approx 1/2 \cdot 12=6$ in three dimensions. To obtain this reduced complexity of $O(N)$ instead of  $O(N^2)$ have been employed in the last decades.

\textbf{Verlet lists}\cite{Verlet1967tables} of interacting particles are set up after a certain number of timesteps after a certain number $m_{\textrm{step}}$ of timesteps via a double loop over all particle pairs: This reduces the prefactor, but not the principal order of the $O(N^2)$ complexity. Another issue is that the particle velocity and the collision rate affects the number of updates. Accordingly, for faster moving
particles and higher densities, a smaller $m_{\textrm{step}}$ is necessary. 

\textbf{Neighborhood tables with linked cell lists}\cite{QuentrecBrot1973} 
are constructed by assigning particles to a grid of cells. The grid size is determined by the particle size or interaction range and the neighborhood algorithm determines the interacting particles in the same cell or neighboring cells, depending on the choice of cell size. A drawback of this method is that in a straightforward, parallelizable implementation, for systems which have empty and high density regions, there is a considerable waste of computer memory for empty neighborhood cells.
(We do not consider the linke list variant of this algorithm\cite{hockney1981computer}, which  is scalar-recursive, constructed from scratch in each timestep and not parallelizable.)

\begin{figure}[h]
\centering
    \includegraphics[clip,width=.6 \hsize]{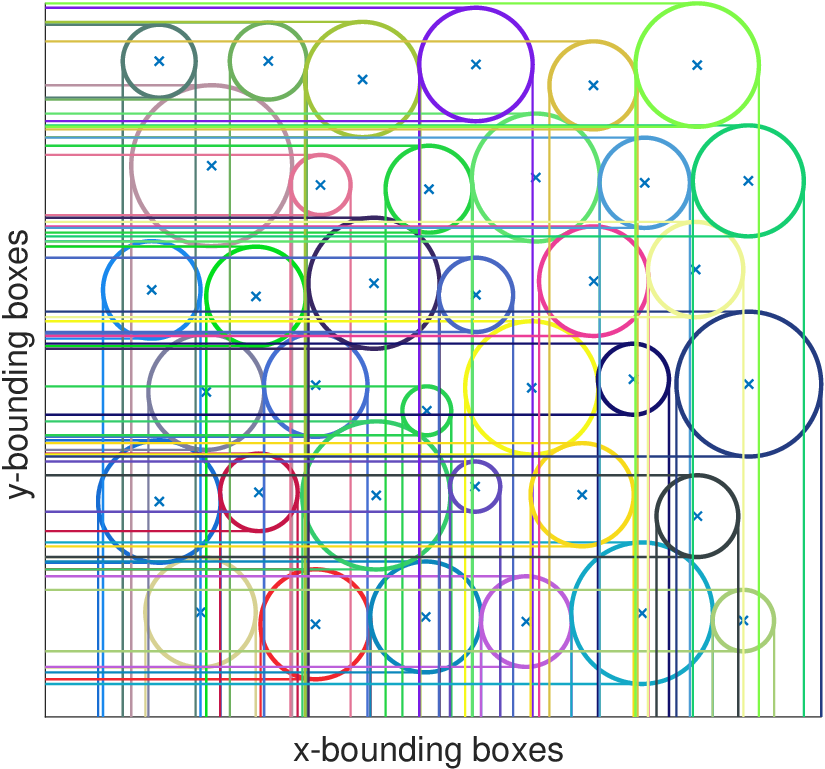}
\caption{Bounding boxes in x- and y-direction must be resorted in the sort-and-sweep approach irrespective if the particles can have contacts along the respectively ``other'' coordinate.}
\label{Fig_noncontact}
\end{figure}

\textbf{Sort and sweep}\cite{BaraffPhDthesis}
uses sorted lists of bounding boxes (upper and lower coordinates for each dimension): When particles change positions, the bounding boxes are updated and re-sorted. Whenever upper coordinates move below lower coordinates, the corresponding particle coordinates are written into a list of possible interactions. The resorting of the (from the previous timestep already partially ordered) list is effectively of $O(N),$ which is practically optimal, and there is no memory overhead, as only the bounding boxes of the particles and the list of interacting particle pairs must be maintained.
The disadvantages of this approach, compared to the previously mentioned methods, are minor, but twofold. On the one hand, coordinates are dealt with independent of each other, so that x-coordinates of bounding boxes must be reordered for particles which cannot interact along the y-axis (respectively the z-coordinates) and vice versa, see Fig.\,\ref{Fig_noncontact}. For this type of algorithm, an efficient parallelization beyond the  inherently possible parallel execution of functions for the  x-, y- and z-coordinates seems to be extremely difficult: 
Because the position changes of the bounding box coordinates in each timestep are relatively minor, incremental and irregularly interspersed over the whole lists, a distribution of the work for a single coordinate on multiple cores does not parallelize well. Even with compiler languages, a more fine-granular parallelization by distributing the resorting of the partially ordered lists gave no speedup\cite{Tenhagen}. It is these drawbacks which we want to overcome by employing tree codes.

\begin{figure}[t]
\centering
\includegraphics[clip,width=\hsize]{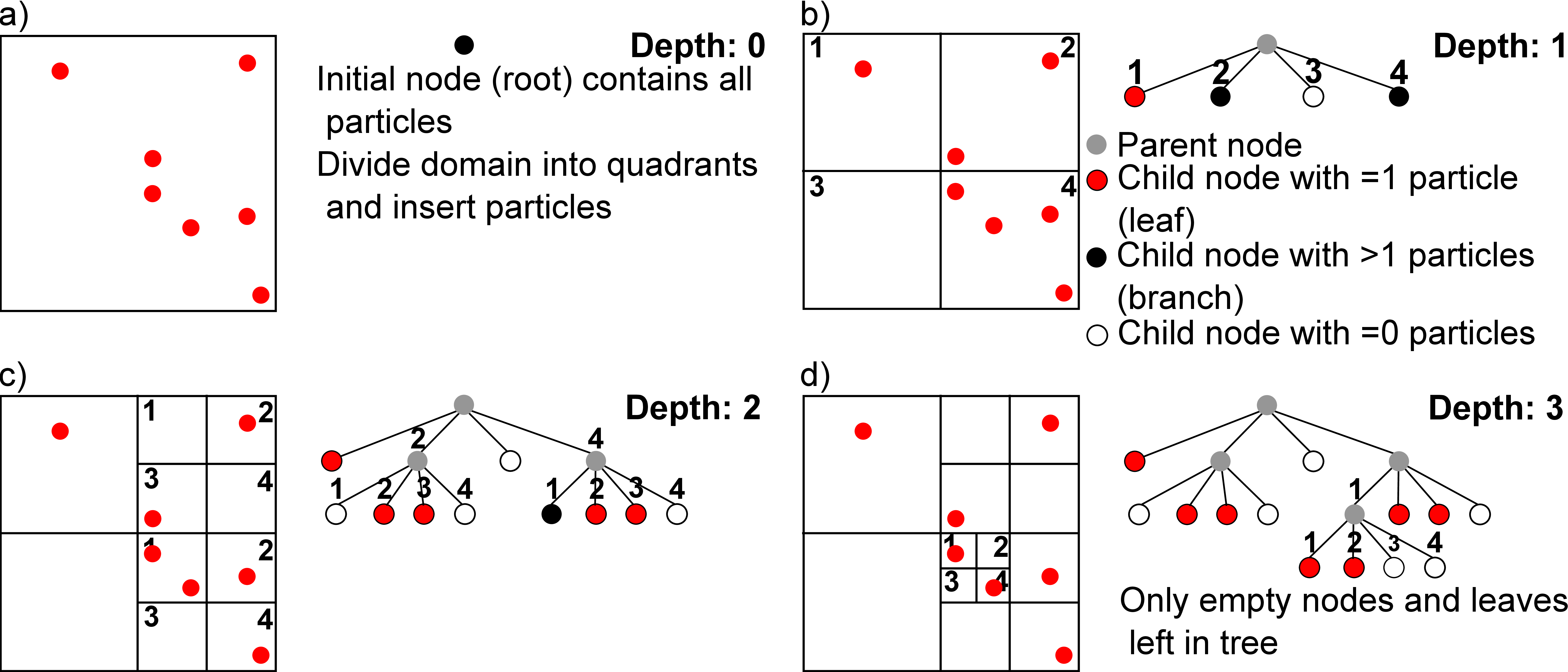} 
\caption{Original particle configuration in a) and successive partitioning and assignment of center of mass positions in ``boxes'' for a given particle configuration in b) to d).}
\label{Fig_partitioning_2Dtree}
\end{figure}

\subsection{Tree codes}
The terminology of ``trees'' was introduced by Cayley in the 19th century and the use of tree data structures in computer science gained momentum from the 1960s onward. The introduction into physics to reduce the computational complexity of the summation of ``long-range'' interactions dates from the 1980s\cite{Appel1985,BarnesHut1986Nature}. Such applications are focused on $\propto 1/r^2$ forces from gravitation, electrostatics  or point vortices. M. C. Lin\cite{Lin93Phd} incorporated  tree codes for her  closest feature algorithms for polytypes and proposed tree codes for neighborhood algorithms in solid modeling (elastic or rigid particles), but did not implement the approach. 

\begin{figure}[h]
\begin{minipage}{.4 \textwidth}
\includegraphics[clip,width=\hsize]{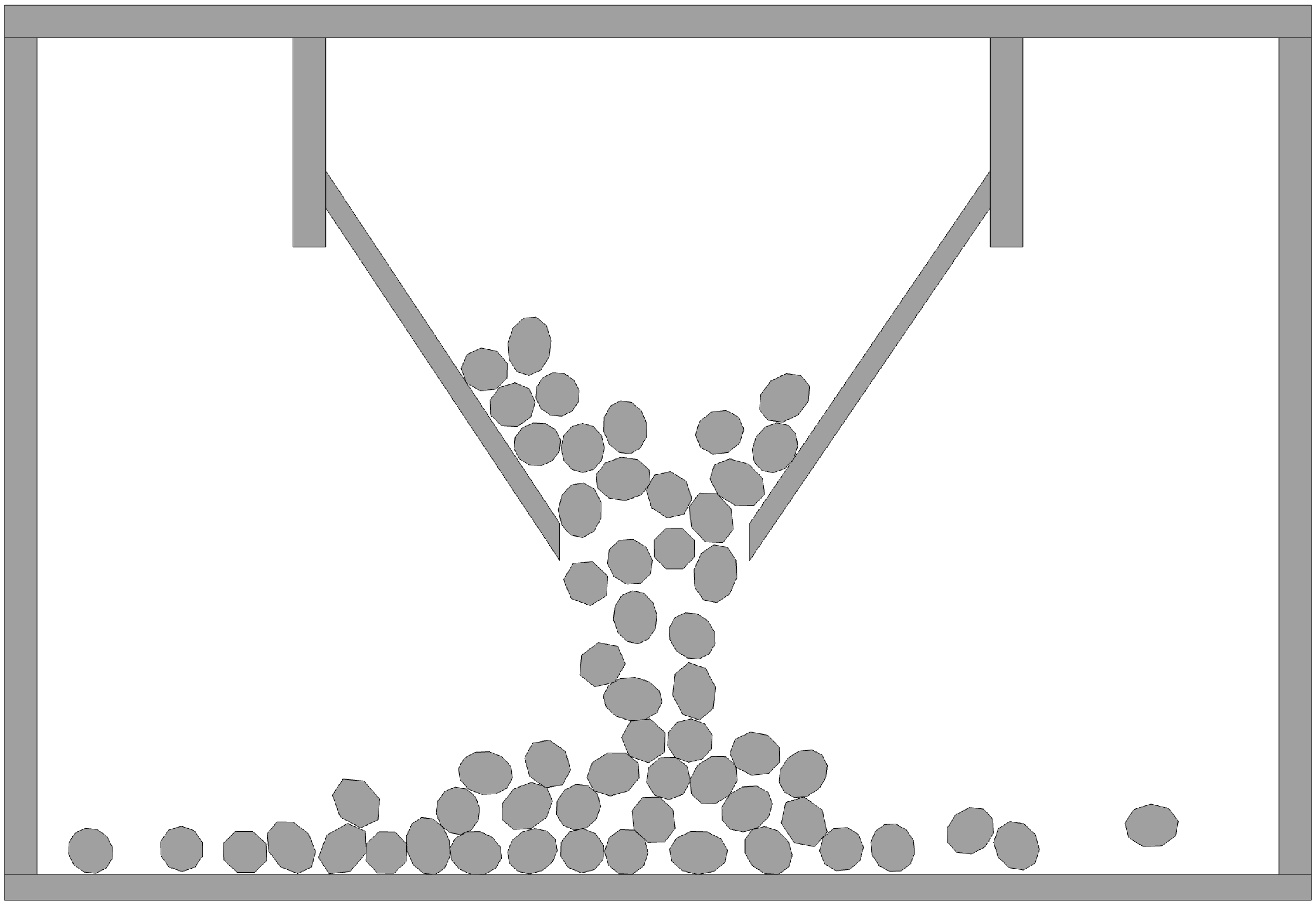} \hfill
\includegraphics[clip,width=\hsize]{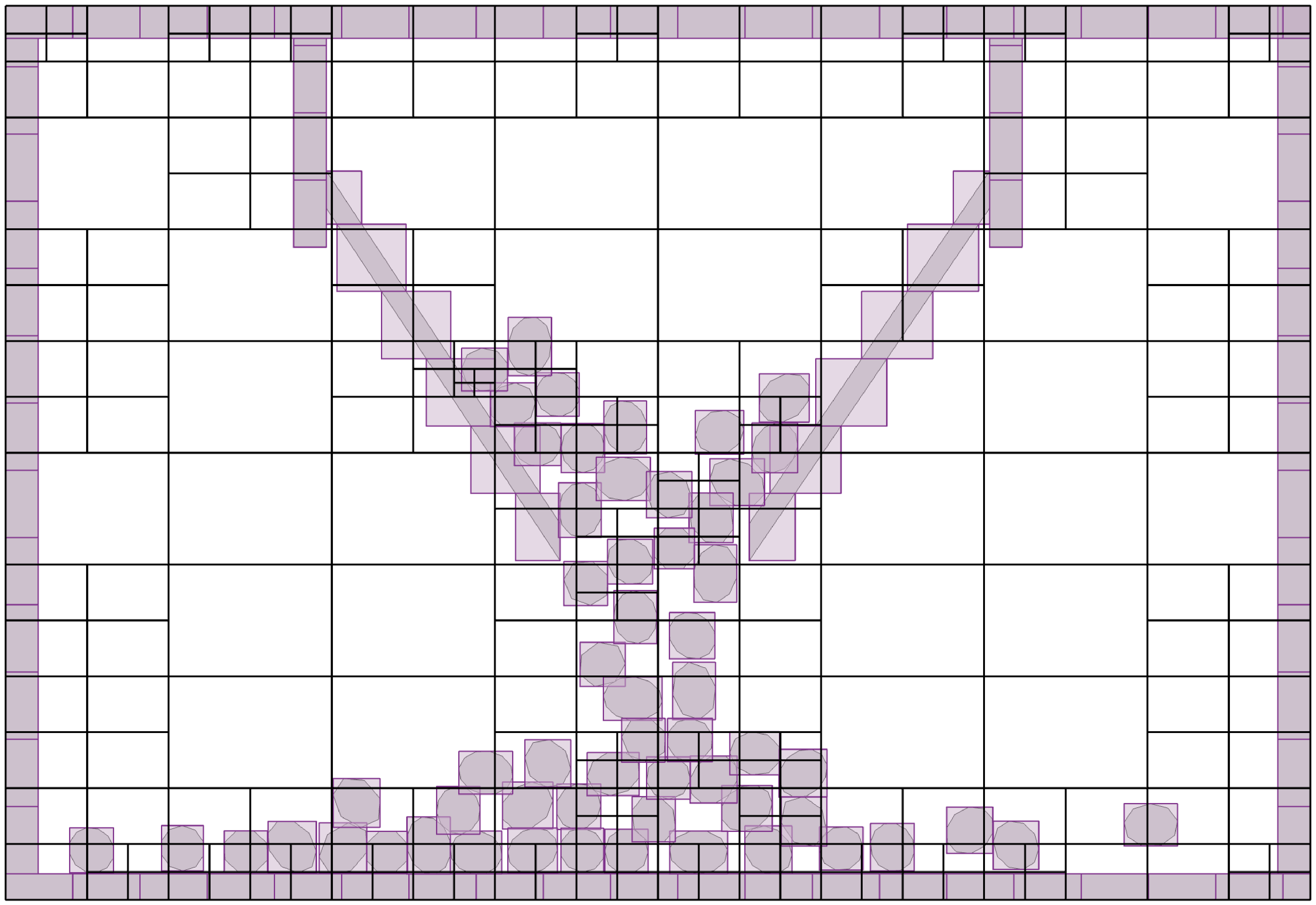}
\end{minipage}
\begin{minipage}{.58 \textwidth}
\includegraphics[clip,width=\hsize]{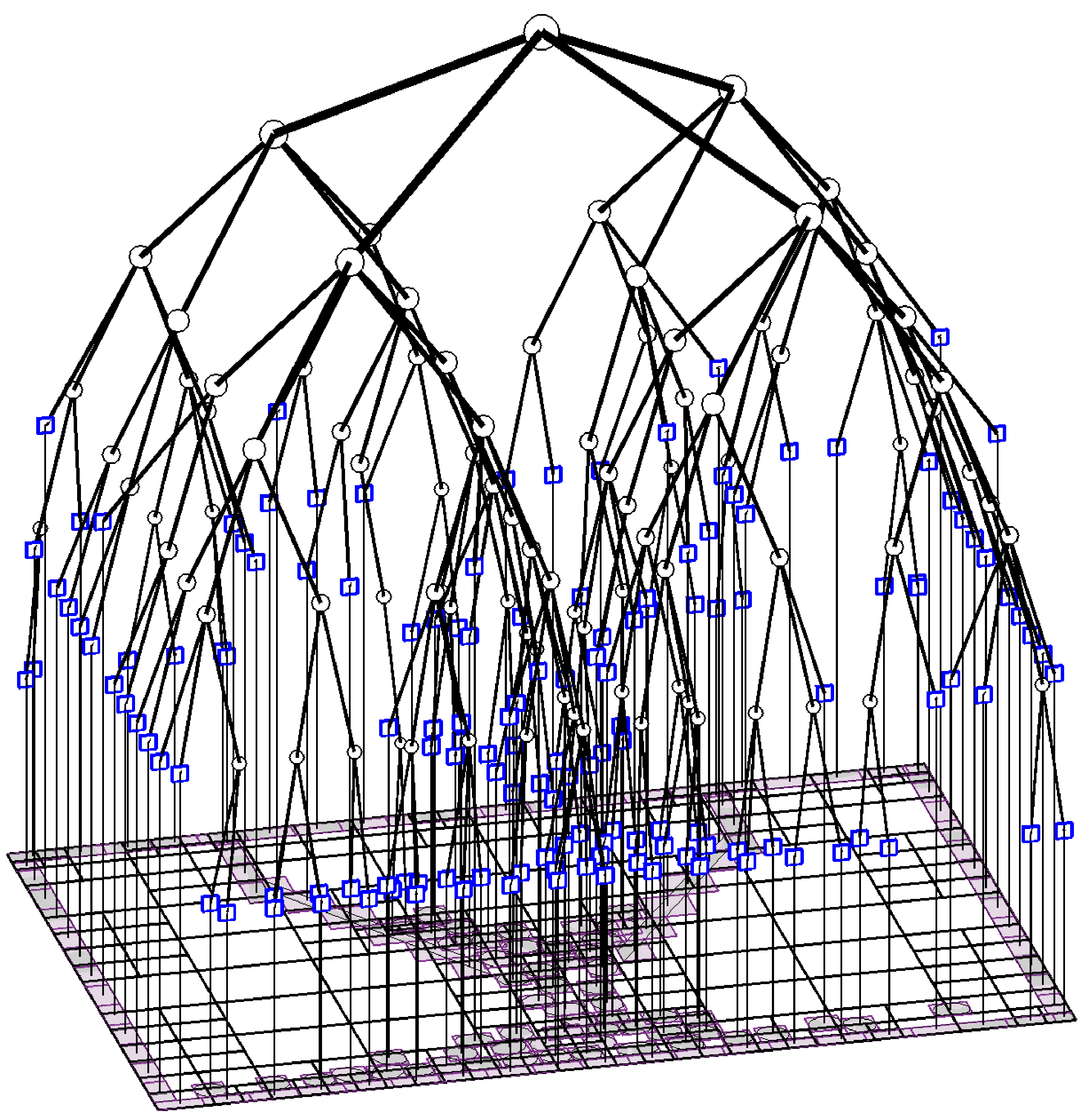}
\end{minipage}
\caption{From particle configuration (left, above) to bounding boxes of the same combination (left, below) to tree structure of the bounding boxes.}
\label{Fig_part_box_tree}
\end{figure}

\section{Tree code implementation}
Tree codes are an application of the divide and conquer principle so that  a one-, two- or three-dimensional region is recursively subdivided into two, four and eight sub-regions. Fig.\,\ref{Fig_partitioning_2Dtree} shows an example for a two-dimensional case:
Starting  from the ``root'', where all particle indices are contained, to the successive partitioning, where in the final configuration of ``leaves'' only a single particle is left in each sub-region. It is obvious that the tree data structure at times needs a much more versatile reconfiguration than the sort and sweep algorithm, which makes only use of a list of bounding boxes which has to be re-sorted.

The aim of this research was to evaluate the efficiency of tree codes for actual DEM simulations relative to  the ``sort and sweep'' neighborhood algorithm. That has a theoretical complexity of $O (N \log N), $ 
which even for a system with $10^6$ particles is $O (10^6 \cdot 6), $ which  for practical purposes is $O(N).$ The actual computational effort at a given timestep goes only in the rearrangement of the bounding boxes and the evaluation of particle bounding boxes where the neighborhood has effectively changed. Because timesteps for DEM simulations are relatively small, the number  of reorderings and actually changed pairs are also relatively small. In comparison, trying to beat sort and sweep when computing neighborhoods for a tree data structure constructed ``from scratch'' in each timestep is hopeless: It would be much more time-consuming, so only  an incremental updating of the existing tree data structure can be competitive performance-wise.

\subsection{What not to do}
In our first, preliminary implementation for one dimension, the benchmarking resulted in a surprising time consumption  of $O(N^2)$ for $N$ particles instead of the expected $O(N \log N).$ The reason was an allocation of an auxiliary vector of length $N$ in each of the $N$ calls for the updating of the tree data structure, which could be avoided by ``recycling'' the same data vector in each function call. This shows how the ease which allows the  fast prototyping via modern programming languages has to be balanced. 
We also will not implement oriented bounding boxes, as this introduces additional complexity in the neighborhood algorithm. We will work with axis-aligned bounding boxes only, to obtain the possible candidates for interaction with minimal computational complexity. Accordingly, the case of narrow separation between particles is deferred to the overlap computation of the DEM algorithm. 

\subsection{Other algorithmic considerations}
 Trees come in two ``flavors'': For binary trees, quadtrees and octrees
(in respectively one, two and three dimensions), the outer boundaries are fixed, while for k-d trees, the dimensions change with changes of the outline of the region. For our type of simulation, the domain boundary can be chosen as fixed, and we will concentrate on the two-dimensional neighborhood algorithm using a quad-tree. We avoided the use of k-d trees, where the structure  depends on the relative and not the absolute positions of the particles, which would complicate the updating. 
The algorithmic issues explained in the following apply also for octrees in three dimensions. There is an issue with the particle size respective their bounding boxes. While tree codes for systems of long-range interactions are written for point particles without any explicit scale, for DEM simulations, a size dispersion has to be taken into account. In particular,  walls are particles which span a significant part of the whole system. This issue is dealt with by subdividing the walls into bounding boxes of the extension of ``average'' particle size, see Fig.\,\ref{Fig_part_box_tree}.
A final difference to systems of point particles with long-range interactions is that instead of partitioned force summations over the system, lists of actually interacting particles must be constructed. 

\begin{figure}[h]
\begin{minipage}{ \textwidth}
\includegraphics[clip,width=\hsize]{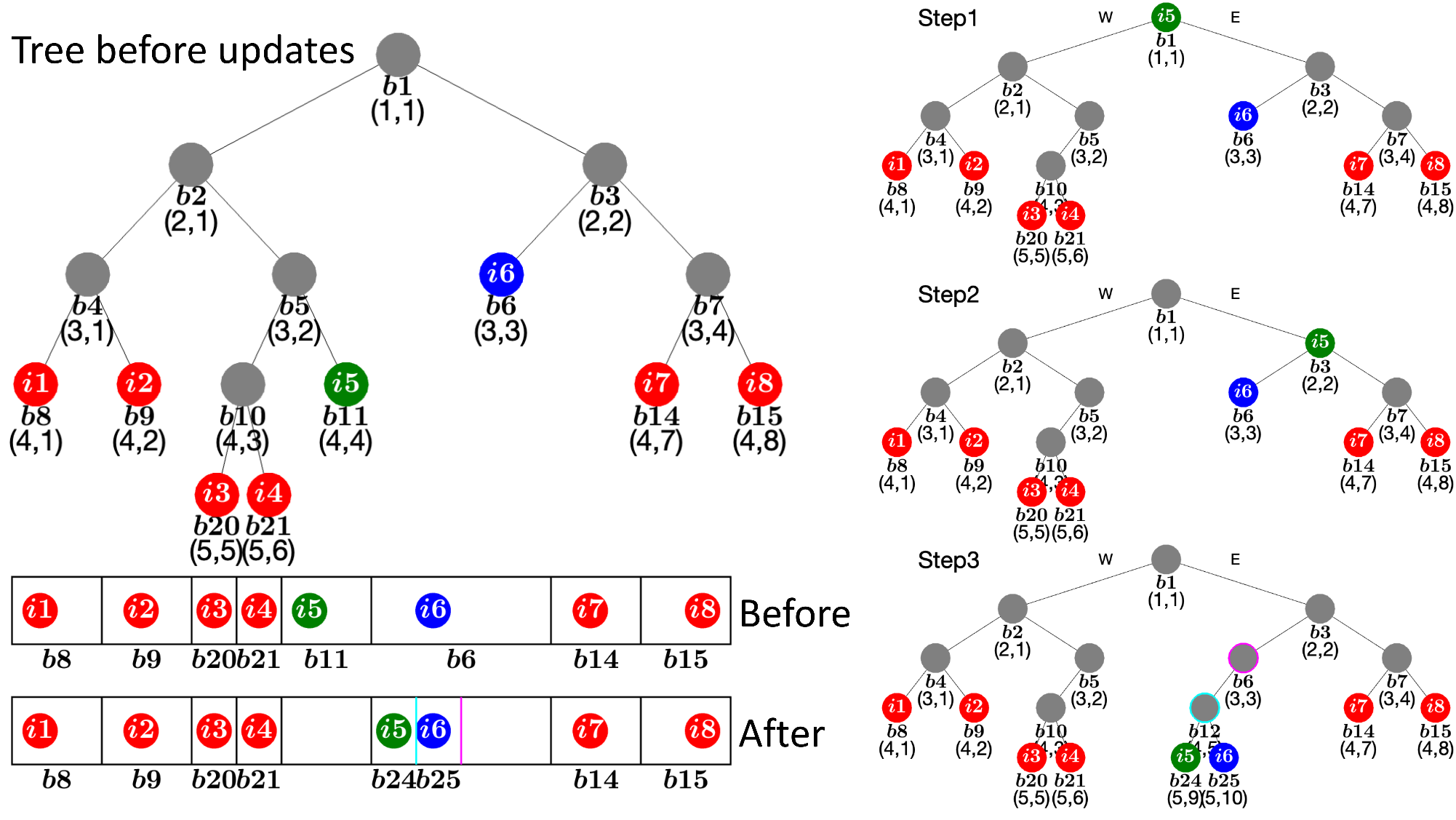} 
\end{minipage}
\caption{Eight particles $i1$ to $i8$ where particle $i5$ has moved to the right so that it must be repositioned in the tree. Because the center of mass of particle $i5$ will end up in the right part of the tree, already in the first step (right column, up) the index $i5$ can be positioned at the root node (on the top, i.e. the apex). From there, $i5$ has to descend, with the decision of the branches according to the relative position of the centers of mass of the other particles. Leaf nodes are drawn in red, parent nodes are drawn in gray, and unoccupied nodes are left empty.}
\label{Fig_tree_update1D}
  \end{figure}

\subsection{Tree updates in one dimension}
The basic principle for updating a three is shown in Fig.\,\ref{Fig_tree_update1D}: After the new  center of mass is updated in the simulation, the particle index is migrated to the new position in the tree based on the center of mass coordinates of the already existing particles. First, it is elevated to the highest possible tree node (in the case of particle i5, the highest position in the tree is b1) based on the computation from the center of mass coordinate without passing through the grey nodes in between from the ``left'' (west) to the ''right'' (east) half of the tree.
From the highest node, it gradually moves downward again according to decision process based on the existing structure and leaf nodes. Fig.\,\ref{Fig_tree_update1D} shows two kinds of nodes: Particle indices pass through the nodes colored gray without any change to the tree structure. Only when a node in another color (red or blue) is reached, the tree structure is changed and the node must be split, or another leaf node be added. While the particle is moving through the grey nodes of the tree, the information about the tree position is stored for each particle, not the other way round (the particle information is not entered in the tree data structure). Accordingly, the rearrangement of the particle indices along the grey nodes of the tree can be done in parallel. Only at the last step, when leave nodes have to be split or added, the tree structure has to be changed. 

\subsection{Tree updates in two dimensions}
\begin{figure}[h]
\centering
\includegraphics[clip,width=.9 \hsize]{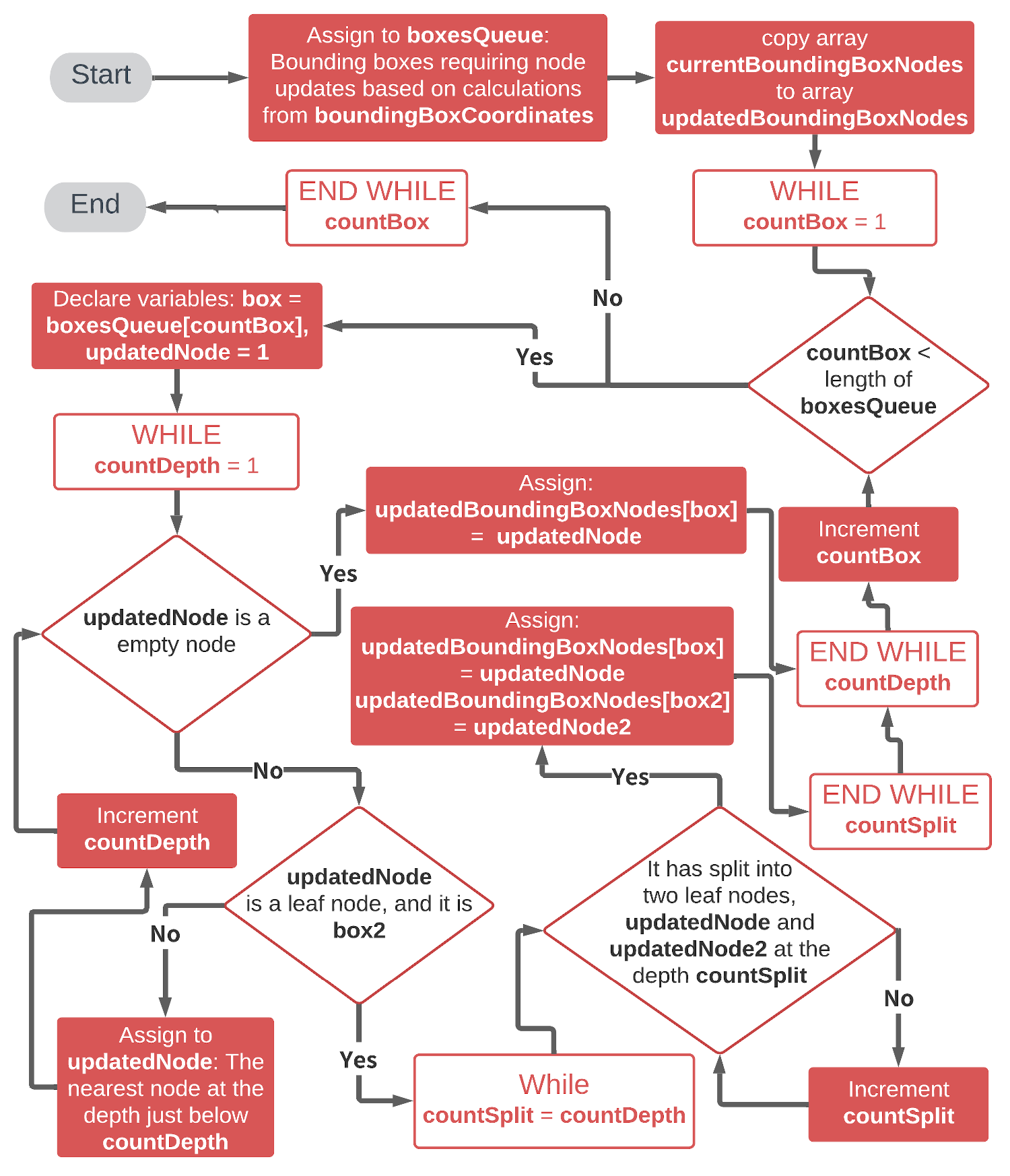} 
\caption{Flow diagram for updating the structure of the two-dimensional neighborhood tree.}
\label{Fig_tree_update2Ddiagram}
\end{figure}

While the principle for the positioning in the new tree is the same for one and two dimensions, decisions and algorithms become more involved due to the additional dimension. Instead of only ``west'' and ``east'' direction, additionally there are ``north'' and ``south''. A flow diagram is shown in Fig.\,\ref{Fig_tree_update2Ddiagram}: 
First, as in one dimension, the position in the discrete boxes is determined. The list of all bounding boxes where changes are necessary is written into \texttt{boxesQueue}. For these bounding boxes, the list of corresponding tree nodes which need changes is written into \texttt{updatedBoundingBoxNodes}.As in one dimension, the particle indices are immediately migrated to the highest node they can reach, based on their coordinates. The remaining algorithm is concerned only with the descent in the tree. Essentially, there are thee possibilities: 
Either, a node can be descended without change, or a single leaf node  must be split. The third possibility, which does not occur in the one-dimensional case is the existence of ``empty'' leaf nodes on the same
level as the lowest occupied leaf nodes. 
The particle indices can directly move into these positions (empty circles in Fig. 2). Accordingly, most of the algorithm in Fig.\,\ref{Fig_tree_update2Ddiagram} consists of bookkeeping which bounding boxes must still be moved and which nodes must be split or transcended. 

  \subsection{Setting up the neighborhood pair list}
Tree-codes can retain the full geometric information about the neighborhood relation of particles and describe not only whether an overlap is possible or not, but also in which direction. First, a list of neighboring nodes \texttt{cellsQueue} (both empty or occupied with bounding boxes) is set up for every bounding box (particle), based on the interaction range, i.e. the size of the largest possible bounding boxes in the region. This list of possible interactions is then reduced based on the actually occupied cells and the size of the bounding boxes. 
In the case of elongated particles, looking only in the next adjacent cell is not sufficient, a wider search range is needed as explained in Fig.\,\ref{Fig_LongerParticleSearch}.

\begin{figure}[h]
\centering
\includegraphics[clip,width=.9 \hsize]{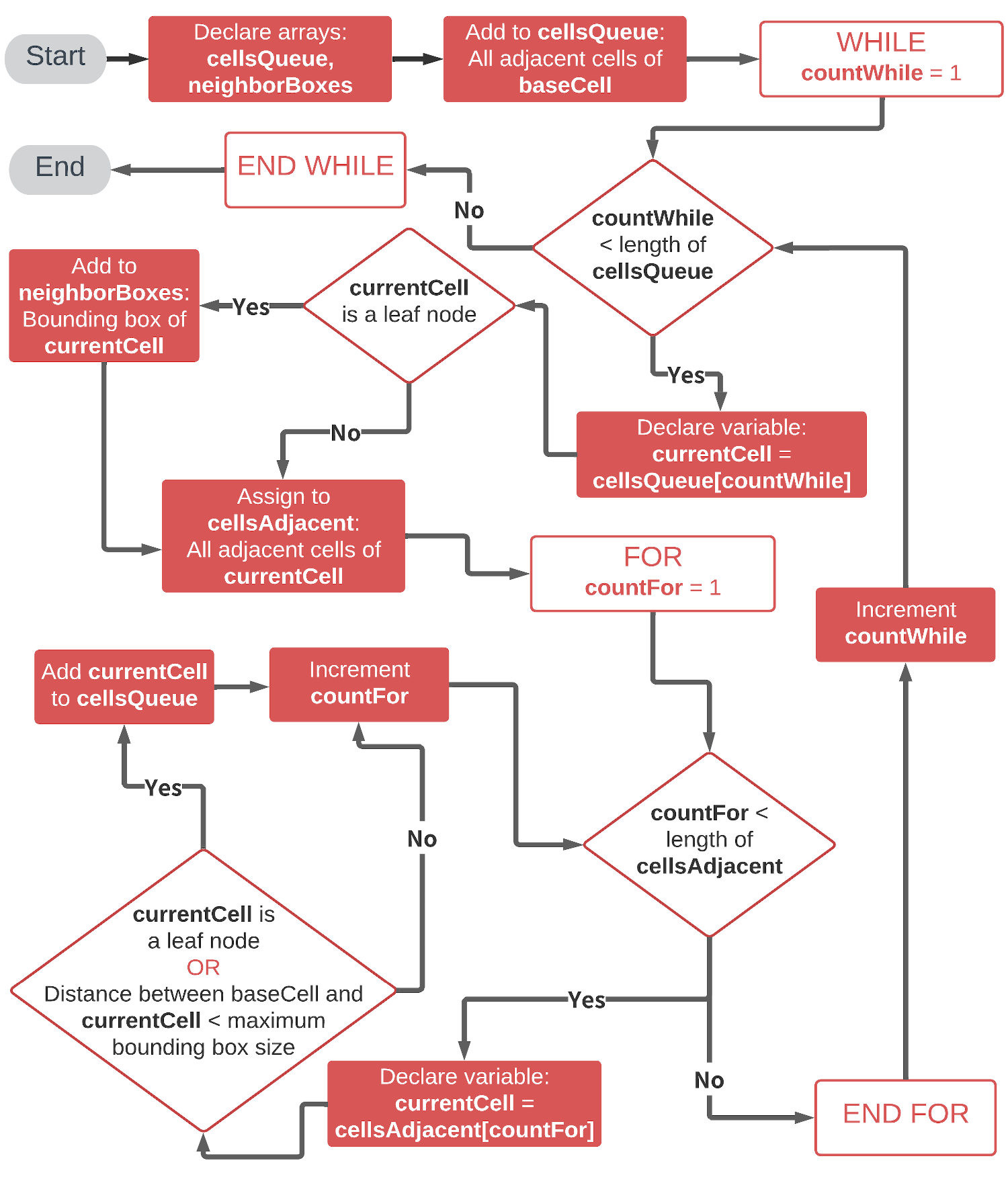} 
\caption{Flow diagram for the neighborhood search inside the two-dimensional neighborhood  tree.}
\label{Fig_neigh_search2Ddiagram}
\end{figure}

\begin{figure}[t]
\centering
\includegraphics[clip,width=\hsize]{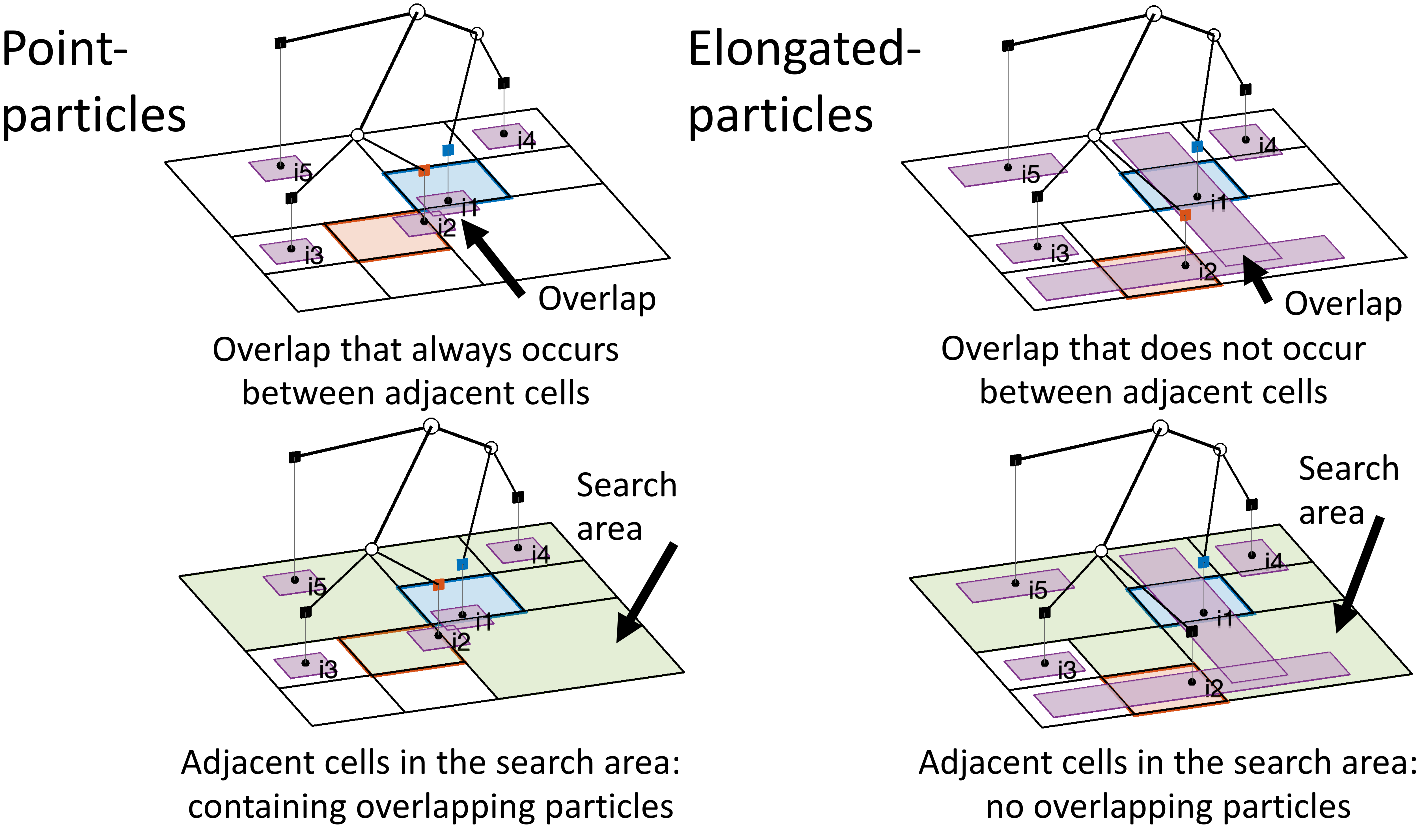} 
\caption{Case where the contact of the non-elongated particle 
in the blue leaf element (left column)  can be identified by searching for the next occupied leaf element while the contact between the elongated particle   in the blue leaf element  (right column) needs a search  beyond the next empty leaf.}
\label{Fig_LongerParticleSearch}
\end{figure}

\section{Timing Comparison}
\begin{figure}[h]
\begin{minipage}{.6 \textwidth}
 \includegraphics[clip,width=\hsize]{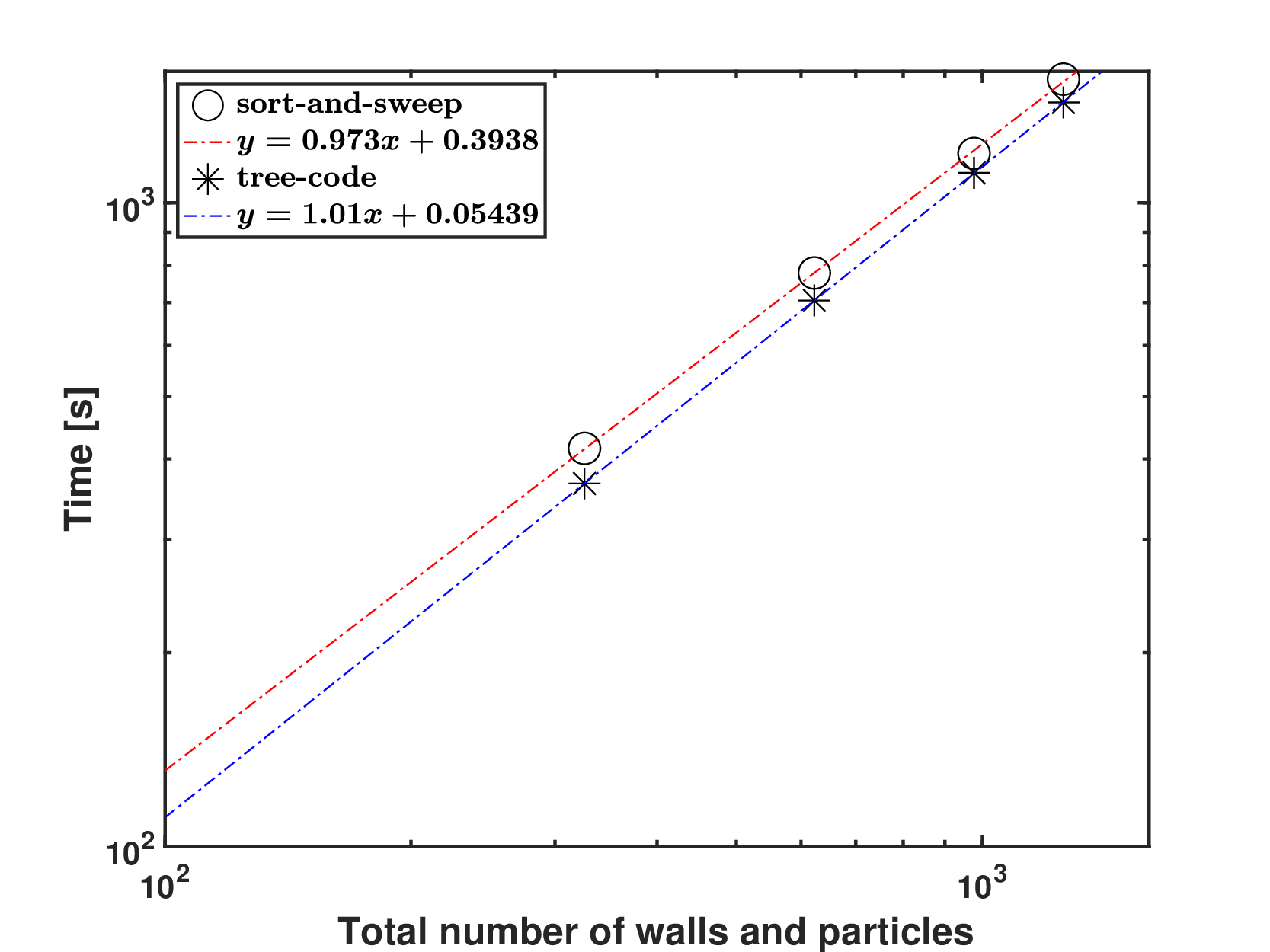} 
\end{minipage} \hfill
\begin{minipage}{.38 \textwidth}
 \includegraphics[clip,width=\hsize]{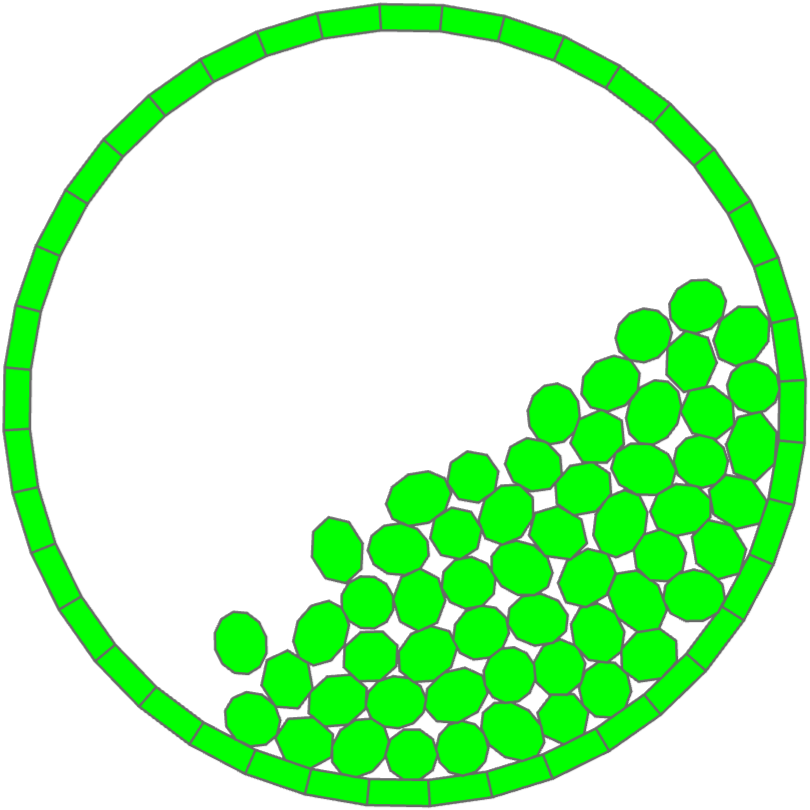} 
\end{minipage} 
\caption{Comparison between the time consumption for sort and sweep and the tree code on the left for the drum on the right. The tree code based neighborhood algorithm is more favorable already for systems of about 180 particles.}
\label{Fig_timingDrum}
\end{figure}

The timings for particles in a  rotating drum Fig.\,\ref{Fig_timingDrum} (right) with different fillings are compared for the neighborhood computation for  sort and sweep and the tree code in  Fig.\,\ref{Fig_timingDrum} (left). The tree code performs obviously better, but that is due to the choice of the system: The change of particle coordinates mentioned in connection with Fig.\,\ref{Fig_noncontact} is enforced by the rotating drum boundary, so that a maximum of boundaries is interchanged, which must be resorted again in the sort and sweep approach. 
 From extrapolation of the lines in Fig.\,\ref{Fig_timingDrum} (left), the breakeven where the tree code will consume about the same time as the sort and sweep algorithm will occur at about 8500 particles for the current implementation. While the tree code also has to update the neighborhoods more often, changes are only necessary in a limited range along the x- and y-direction, while for the sort and sweep algorithm, the whole coordinate range is globally affected. In contrast, for system with relatively few changes of the coordinates as in hopper flow geometries like Fig.\,\ref{Fig_part_box_tree}, the search and sweep algorithm performs better, but the time consumption is mostly comparable.

\section*{Conclusions and Summary}
We have successfully implemented a neighborhood algorithm based on the updating of quadtrees for discrete element simulations in two dimensions. The principles are also applicable to octrees in three dimensions. While the performance of the sorting part of the tree code is competitive with the sort and sweep algorithm, there are better possibilities for parallelizing the construction of the neighborhood list.

\bibliographystyle{unsrt}
\bibliography{TreecodesNeigh}

\begin{thebibliography}{1}

\bibitem{Verlet1967tables}
Loup {Verlet}.
\newblock {Computer ``Experiments'' on Classical Fluids. I. Thermodynamical
  Properties of Lennard-Jones Molecules}.
\newblock {\em Physical Review}, 159(1):98--103, July 1967.

\bibitem{QuentrecBrot1973}
B.~Quentrec and C.~Brot.
\newblock New method for searching for neighbors in molecular dynamics
  computations.
\newblock {\em J. Comput. Phys.}, 13:430--432, 1973.

\bibitem{hockney1981computer}
R.W. Hockney and J.W. Eastwood.
\newblock {\em Computer Simulation Using Particles}.
\newblock McGraw-Hill International Book Company, 1981.

\bibitem{BaraffPhDthesis}
D.~Baraff.
\newblock {\em Dynamic Simulation of Non-Penetrating Rigid Bodies}.
\newblock PhD thesis, Cornell University, 1992.

\bibitem{Tenhagen}
Robin Tenhagen.
\newblock Shared-memory parallelization of the sort-and-sweep algorithm via
  openmp.
\newblock Research report of the {JUSST}-program of the {U}niversity of
  {E}lectro-{C}ommunications, 2013.

\bibitem{Appel1985}
Andrew~W. Appel.
\newblock An efficient program for many-body simulation.
\newblock {\em SIAM Journal on Scientific and Statistical Computing},
  6(1):85--103, 1985.

\bibitem{BarnesHut1986Nature}
Josh {Barnes} and Piet {Hut}.
\newblock {A hierarchical O(N log N) force-calculation algorithm}.
\newblock {\em Nature}, 324(6096):446--449, dec 1986.

\bibitem{Lin93Phd}
Ming~Chieh Lin.
\newblock {\em Efficient Collision Detection for Animation and Robotics}.
\newblock PhD thesis, University of California, Berkeley, Berkeley, CA, 1993.

\end{thebibliography}
\end{document}